\documentclass[aps,prb,amsmath,amssymb,amsfont,showpacs,superscriptaddress,reprint]{revtex4-1}
\usepackage{amsmath}
\usepackage{bm}
\usepackage[dvipsnames]{xcolor}
\usepackage{braket}
\usepackage[artemisia]{textgreek}  
\usepackage{float}
\usepackage{graphicx}      
\usepackage{siunitx}
\usepackage{wasysym}

\begin{document}

% Title of the article
\title{Optical-microwave pump-probe studies of electronic properties in novel materials}
%\titlerunning{Optical-microwave pump-probe studies of electronic properties in novel materials}

% Authors

\author{S. Kollarics}
\affiliation{Department of Physics, Budapest University of Technology and Economics and MTA-BME Lend\"{u}let Spintronics Research Group (PROSPIN), P.O. Box 91, H-1521 Budapest, Hungary}
\affiliation{Laboratory of Physics of Complex Matter, \'{E}cole Polytechnique F\'{e}d\'{e}rale de Lausanne, Lausanne CH-1015, Switzerland}

\author{A. Bojtor}
\affiliation{Department of Physics, Budapest University of Technology and Economics and MTA-BME Lend\"{u}let Spintronics Research Group (PROSPIN), P.O. Box 91, H-1521 Budapest, Hungary}

\author{K. Koltai}
\affiliation{Department of Physics, Budapest University of Technology and Economics and MTA-BME Lend\"{u}let Spintronics Research Group (PROSPIN), P.O. Box 91, H-1521 Budapest, Hungary}

\author{B. G. M\'{a}rkus}
\affiliation{Department of Physics, Budapest University of Technology and Economics and MTA-BME Lend\"{u}let Spintronics Research Group (PROSPIN), P.O. Box 91, H-1521 Budapest, Hungary}
\affiliation{Laboratory of Physics of Complex Matter, \'{E}cole Polytechnique F\'{e}d\'{e}rale de Lausanne, Lausanne CH-1015, Switzerland}
\author{K. Holczer}
\affiliation{Department of Physics and Astronomy, UCLA, Los Angeles, California 90095-1547, United States}

\author{J. Volk}
\affiliation{Nanosensors Research Group, Centre for Energy Research, Budapest, Hungary}

\author{G. Klujber}
\affiliation{Institute of Nuclear Techniques, Budapest University of Technology and Economics, M\H{u}egyetem rkp. 9, H-1111 Budapest, Hungary}

\author{M. Szieberth}
\affiliation{Institute of Nuclear Techniques, Budapest University of Technology and Economics, M\H{u}egyetem rkp. 9, H-1111 Budapest, Hungary}

\author{F. Simon}
\affiliation{Department of Physics, Budapest University of Technology and Economics and MTA-BME Lend\"{u}let Spintronics Research Group (PROSPIN), P.O. Box 91, H-1521 Budapest, Hungary}
\affiliation{Laboratory of Physics of Complex Matter, \'{E}cole Polytechnique F\'{e}d\'{e}rale de Lausanne, Lausanne CH-1015, Switzerland}

%\author{
%  S. Kollarics\textsuperscript{\textsf{\bfseries 1,5}},
%  A. Bojtor\textsuperscript{\textsf{\bfseries 1}},
%  K. Koltai\textsuperscript{\textsf{\bfseries 1}},
%  B. G. M\'{a}rkus\textsuperscript{\textsf{\bfseries 1,5}},
%  K. Holczer\textsuperscript{\textsf{\bfseries 2}},
%  J. Volk\textsuperscript{\textsf{\bfseries 3}},
%  G. Klujber\textsuperscript{\textsf{\bfseries 4}},
%  M. Szieberth\textsuperscript{\textsf{\bfseries 4}},
%  F. Simon\textsuperscript{\Ast, \textsf{\bfseries 1,5}}
%  }

%\authorrunning{S. Kollarics \emph{et al.}}

%E-mail-address of corresponding author
%\mail{e-mail: \textsf{f.simon@eik.bme.hu}, Phone: +36-1-463-1215, Fax: +36-1-463-4180}

% author's affiliations/addresses
%\institute{%
%  \textsuperscript{1}\,Department of Physics, Budapest University of Technology and Economics and MTA-BME Lend\"{u}let Spintronics Research Group (PROSPIN), P.O. Box 91, H-1521 Budapest, Hungary\\
%  \textsuperscript{2}\,Department of Physics and Astronomy, UCLA, Los Angeles, California 90095-1547, United States\\
%  \textsuperscript{3}\,Nanosensors Research Group, Centre for Energy Research, Budapest, Hungary\\
%  \textsuperscript{4}\,Institute of Nuclear Techniques, Budapest University of Technology and Economics, M\H{u}egyetem rkp. 9, H-1111 Budapest, Hungary\\
%  \textsuperscript{5}\,Laboratory of Physics of Complex Matter, \'{E}cole Polytechnique F\'{e}d\'{e}rale de Lausanne, Lausanne CH-1015, Switzerland\\
%  }

% Please select about four verbal keywords for your manuscript.

\begin{abstract}Combined microwave-optical pump-probe methods are emerging to study the quantum state of spin qubit centers and the charge dynamics in semiconductors. A major hindrance is the limited bandwidth of microwave irradiation/detection circuitry which could be overcome with the use of broadband coplanar waveguides (CPW). 
We present the development and performance characterization of two spectrometers: an optically detected magnetic resonance spectrometer (ODMR) and a microwave detected photoconductivity measurement. In the first method light serves as detection and microwaves excite the investigated medium, while in the second the roles are interchanged. The performance is demonstrated by measuring ODMR maps on the nitrogen-vacancy center in diamond and time resolved photoconductivity in \textit{p}-doped silicon. The results demonstrate both an efficient coupling of the microwave irradiation to the samples as well as an excellent sensitivity for minute changes in sample conductivity.\end{abstract}

\maketitle

\section{Introduction}
Optical pump-probe spectroscopy and microscopy experiments represent an important branch of tools to study the conduction dynamics in novel materials \cite{PumpProbeRSI} or non-equilibrium states \cite{GedikRSI}. Alternatively, electromagnetic radiation with very different wavelength can be combined in a similar pump-probe way, which enables to study physical phenomena occurring on the different energy scales. The combined microwave-optical pump probe methods enable for example to detect microwave induced spin transitions with optics, a method known as optically detected magnetic resonance (ODMR). The reverse situation occurs in microwave detected photoconductivity ($\mu$-PCD), where optically induced non-equilibrium charge carriers cause a change in the reflected microwave intensity, thus enabling a study of charge recombination in a time resolved manner. 
ODMR \cite{odmrbook} spectroscopy is a powerful tool to study the spin states of novel systems such as carbon nanotubes \cite{Stich}, fullerenes \cite{Lane} and nitrogen-vacancy centers in diamond \cite{Gruber} for applications in spintronics \cite{Neumann}, quantum computing \cite{Harneit} or light emitting diodes \cite{Shinar}. This technique benefits from the high energy resolution of the microwaves (down to a few 100 kHz, i.e. neV energy range) with the high efficiency of optical {\color{black}photon} detection.  It is even possible to detect the change in luminescence of a single molecule \cite{Wrachtrup93}. 
Microwave detected photoconductivity decay measurement ($\mu$-PCD) \cite{mupcd} is widely used in semiconductor industry to study impurity concentration in silicon wafers \cite{Berger} or non-silicon semiconductors such as CdTe \cite{Novikov2010}. This contactless method enables manufacturers to determine impurity concentrations with great accuracy and researchers to gain information on charge carrier dynamics on the nanosecond timescale. 
A key element of the instruments in both cases is the microwave irradiation unit, which is most often a resonator, offering a high microwave power to electric or magnetic field conversion due to the resonator quality factor, at the cost of a reduced bandwidth. Alternatively, microwave antennas \cite{antenna} are often used but these have usually smaller bandwidth and worse microwave power to magnetic field conversion ratio. 
In principle compact, non-resonant waveguides combine both the large bandwidth with the good filling factor, which results in an excellent microwave power to magnetic field conversion. Coplanar waveguides \cite{cpwarticle,broadbandcpw} and slot lines \cite{slotline} consist of metallic lines placed on high dielectric constant substrates. This layout confines the microwave fields traveling in a quasi-TEM mode \cite{cpwbook} to a small volume giving a huge electromagnetic energy density competing with classical resonators. These miniaturized waveguides can be combined with surface mount devices (SMDs) \cite{rfbook} and other circuit elements. CPWs also opened a way towards microwave device miniaturization such as production of millimeter-scale circulators \cite{circulator}. In case of resonators, the access to the sample such as optical illumination or application of external electric field is limited and it affects the performance of the resonator, e.g. a window on the wall of the resonator lowers the $Q$-factor. The broadband operation of CPWs makes frequency swept experiments such as permeability measurement \cite{permeability} and ferromagnetic resonance studies \cite{cpwfmr} also very effective.
Herein, we present the development and performance characterization of two instruments, which both operate with coplanar waveguides. We present a broadband optically detected magnetic resonance spectrometer which allowed to study quantum states of nitrogen-vacancy centers in diamond. The microwave detected photoconductivity decay instrument has a good sensitivity for small changes in sample conductivity during the transient decay of the charge carriers. 

\section{The spectrometer setups}
\subsection{The ODMR spectrometer}
\begin{figure}[!h]
	\centering
	\includegraphics*[width=\linewidth]{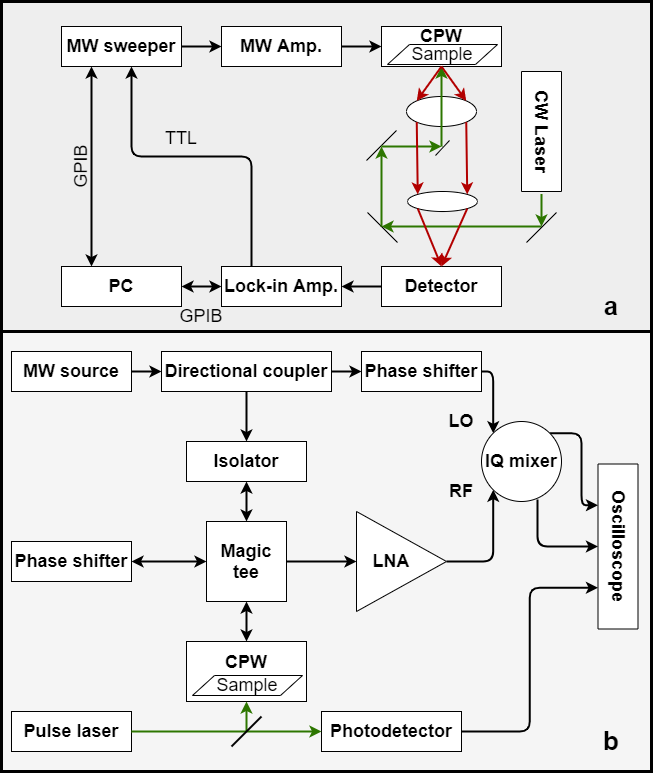}
	\caption{\emph{Block diagram of the optically detected magnetic resonance setup (a) and the microwave detected photoconductivity decay setup (b). Passive attenuators and DC blocks are not shown to retain clarity.}}
	\label{draw}
\end{figure}

The setup of the optically detected magnetic resonance spectrometer is shown in Fig. \ref{draw}. The coplanar waveguide holding the sample is illuminated using a continuous wave laser ($532\ \text{nm}$ Coherent Verdi 5G frequency doubled Nd:YAG). The luminescent light is collected and focused by the so called "image relay" consisting of an achromatic doublet pair (Thorlabs MAP1030100-B) to a spectrograph (Horiba JY iHR320) equipped with a photomultiplier tube (Hamamatsu R2658P) and a single pixel InGaAs detector (Horiba DSS-IGA010L). The former covers the ultraviolet (UV) to near infrared (NIR) range ($185-1010\ \text{nm}$), while the latter is optimized for the NIR range ($1000-1900\ \text{nm}$). The detector output is connected to a lock-in amplifier (Stanford Research Systems, SR830) through a transimpedance amplifier. The output of the microwave signal generator (HP83751B) is chopped by the TTL signal coming from the lock-in amplifier. The microwaves are fed into an amplifier (Kuhne KuPa 270330-10A, $\text{Gain}=40\ \text{dB}$) and then sent to the CPW which has a $50\ \mathrm{\Omega}$ termination at its end. The CPW is placed on a two-axis miniature dovetail translation stage (Thorlabs DT12XY/M) during room temperature measurements. To apply external magnetic field, the stage and the other optics parts are placed on a rail that can be rolled into the middle of an electromagnet with a magnetic field up to 1.2 T.

\subsection{The microwave detected photoconductivity decay spectrometer}

Fig. \ref{draw} shows the microwave photoconductivity decay measurement setup which is a modification of our previous resonator-based, time resolved $\mu$-PCD setup described elsewhere \cite{Gyure}. A Q-switch pulsed laser ($527\ \text{nm}$ Coherent Evolution-15, frequency doubled Nd:YLF) with $1\ \text{kHz}$ repetition rate{\color{black}, pulse energy of $150\ \mu\text{J}$ and pulse width of $1.7\ \mu\text{s}$} is used for excitation of charge carriers in the sample placed on the CPW. 

The microwave oscillator (Kuhne MKU LO 8-13 PLL) serves as the source of the probing MW field and as the local oscillator of a double-balanced IQ mixer. The directional coupler sends half of the incoming power through an isolator (to avoid damage of the oscillator unit by reflected power) to the magic tee, the other half drives the IQ mixer. Half of the incoming microwave power is directed towards the CPW, the other half to the port with a phaseshifter and a variable attenuator attached. The waves reflected from the CPW and the other port interfere with each other on the 4th port allowing us to get rid off the DC reflection by setting the attenuation and phase properly. The previously described MW bridge is followed by a low noise amplifier (JaniLab Inc. $\text{Gain}=15\ \text{dB}$) and the amplified signal is fed to the IQ mixer. The I (in-phase) and Q (quadrature) signals downconverted from the detected RF signal are digitized with an oscilloscope (Tektronix MDO-3024, $\text{BW}=200\ \text{MHz}$) which is triggered by the signal coming from the photodetector (Thorlabs DET36A/M) sensing the laser pulse.

\subsection{The cryostat}
To carry out cryogenic measurement we built a closed cylindrical compartment with optical access through a quartz window ($\diameter$1") and MW connection using an UHV SMA connector. The CPW is fixed to a copper cube attached to the inner tube that can be filled with liquid nitrogen from the outside. The space between the two tubes where the CPW is placed is evacuated using a turbomolecular pump (Pfeiffer HiCube Eco). Vacuum level of $10^{-6}\ \text{mbar}$ is easily achievable and grants long enough measurement time without any condensation problems on the outside or heating of the sample.

\subsection{Samples}
ODMR measurements were performed on single crystal ($3\times 3\times 0.3\ \text{mm}$) Type-Ib diamond samples (Element Six Ltd.) produced by high pressure high temperature method (HPHT), containing less than $200\ \text{ppm}$ substitutional nitrogen. After neutron irradiation of the diamond  plates in the Training Reactor {\color{black}of the Institute of Nuclear Techniques (located at the Budapest University of Technology and Economics)} for 8 hours at 100 kW power with total fluence of appx. $10^{17}\ 1/\text{cm}^2$ ($3\cdot 10^{16}\ 1/\text{cm}^2$ in the $100\ \text{eV}$ - $1\ \text{MeV}$ range), the  samples were annealed under dynamic vacuum ($10^{-5}\ \text{mbar}$) at $800-\SI{1000}{\celsius}$ to help the diffusion of vacancies creating the nitrogen-vacancy centers. To dissolve surface contamination, samples were put into vials containing a mixture of acetone and isopropyl-alcohol and bath-sonicated for 15 minutes.

Microwave detected photoconductivity decay measurements were carried out on phosphorus doped silicon wafer with an approximate surface of $10\ \text{mm}^2$ and thickness of $200\ \mu\text{m}$. The resistivity of the sample was $0.528\ \mathrm{\Omega}\cdot$cm, as determined by a four-point sheet resistivity measurement technique. 

\section{Spectrometer characterization}

\subsection{ODMR on NV centers in diamond}

\begin{figure}[!h]
	\centering
	\includegraphics*[width=\linewidth]{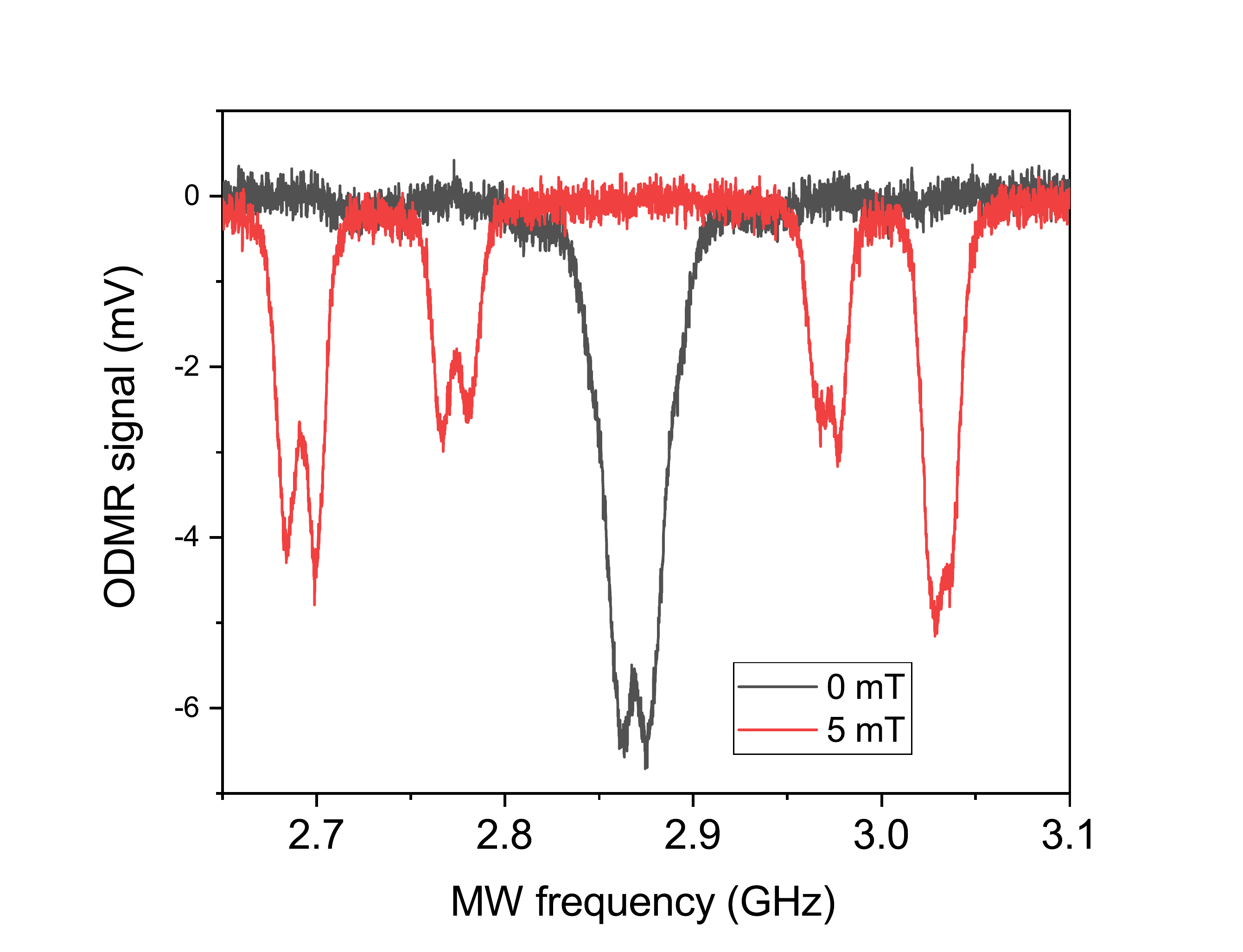}
	\caption{\emph{Optically detected magnetic resonance on nitrogen-vacancy centers in diamond {\color{black} detected at $680\ \text{nm}$}. In zero magnetic field (black line) the resonances are degenerate while applying external magnetic field of $5\ \text{mT}$ (red line) resolves the degeneracy.}}
	\label{fig_odmr}
\end{figure}

Here, we present a range of measurements to demonstrate the versatile applicability of coplanar waveguides. First, ODMR measurements in variable external magnetic fields are shown along with a wavelength resolved zero field ODMR map. We carried out ODMR measurements on diamond samples to demonstrate the advantages of applying CPWs for MW irradiation. Fig. \ref{fig_odmr} shows microwave frequency sweeps performed in zero magnetic field and also in $5\ \text{mT}$.

\begin{figure}[!h]
	\centering
	\includegraphics*[width=\linewidth]{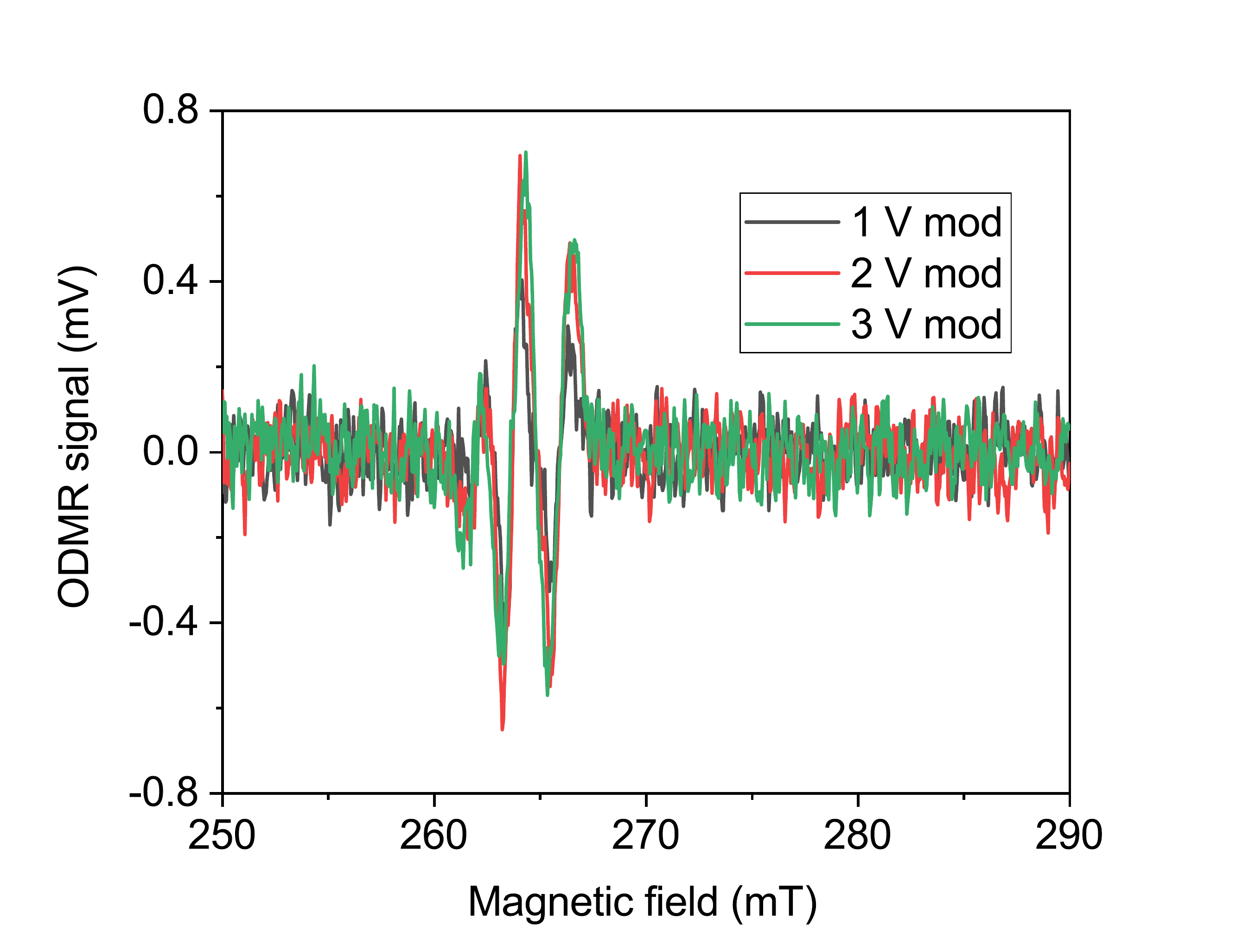}
	\caption{\emph{Frequency modulated and magnetic field swept ODMR measurement at $9.2\ \text{GHz}$ on nitrogen-vacancy centers in diamond {\color{black} detected at $680\ \text{nm}$}. Note that the derivative Lorentzian lineshapes are due to the frequency modulation similar to magnetic field modulation found in conventional electron spin resonance spectrometers.}}
	\label{fig_fm}
\end{figure}

These spectra were acquired in a constant external magnetic field while the microwave frequency was swept and chopped by the TTL signal of the lock-in amplifier. Fig. \ref{fig_fm} on the other hand shows a magnetic field sweep at constant microwave frequency of $9.2\ \text{GHz}$. Here the MW amplifier was replaced with one working in the X-band (Kuhne KuPa 9001250-2A, $\text{Gain}=26\ \text{dB}$). This detection method is essentially the same as if it was a conventional electron spin resonance but the signal is detected optically. The principle of detection is that the microwave field is frequency modulated using the lock-in amplifiers local oscillator as a sine source connected to the microwave oscillator. In Fig. \ref{fig_fm}. the intensity of such ODMR signal is examined as the amplitude of the modulating sine wave increases. The modulation depth was set to $6\ \text{MHz/V}$ meaning a maximum deviation of $18\ \text{MHz}$ in this measurement, however, above $2\ \text{V}$ no significant increase in the ODMR amplitude can be seen.

\begin{figure}[!h]
	\centering
	\includegraphics*[width=\linewidth]{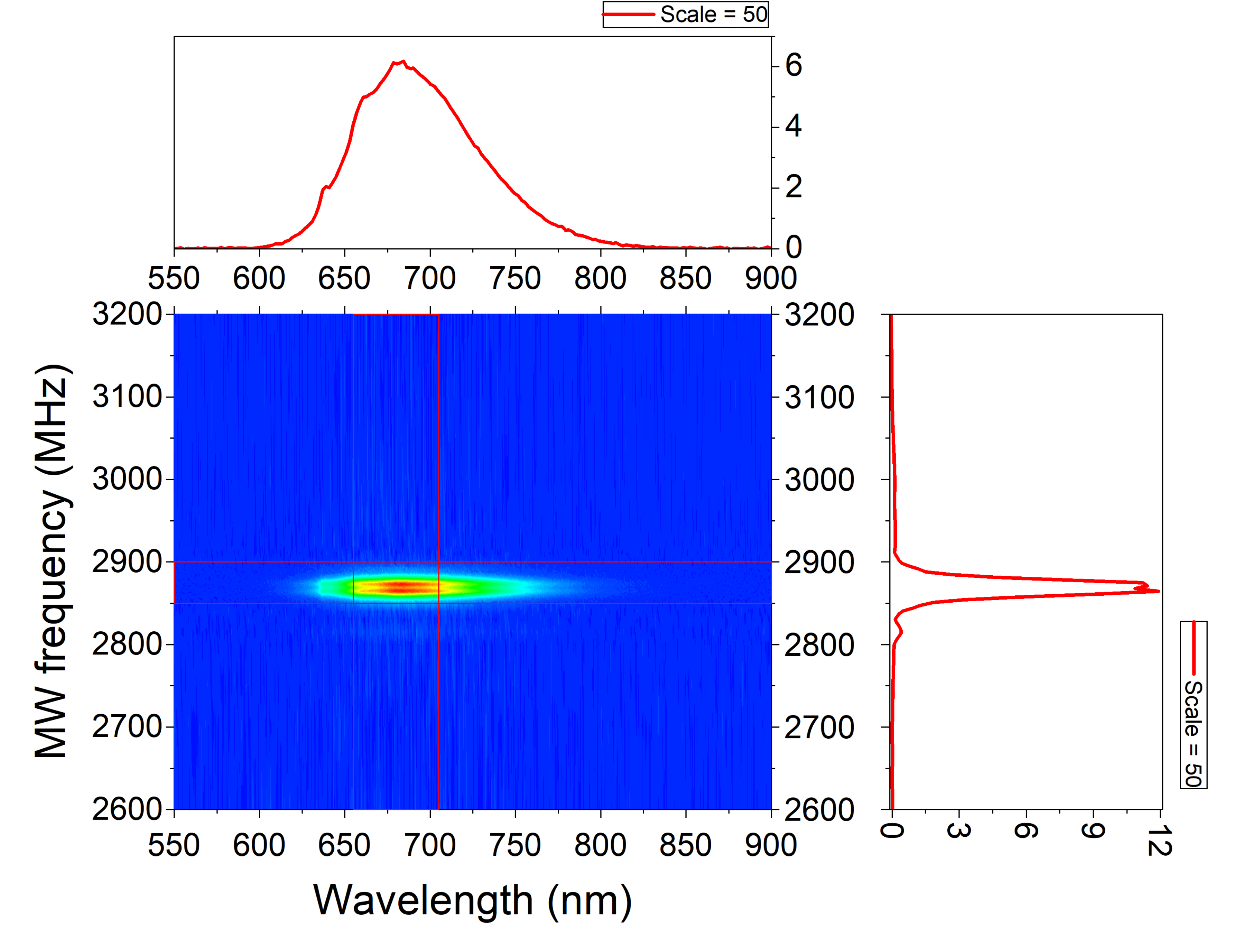}
	\caption{\emph{Optically detected magnetic resonance map. The upper panel is the wavelength resolved microwave frequency averaged {\color{black}(taken in the range $2849-2899\ \text{MHz}$)} photoluminescence similar to the well-known PL spectrum of NV centers. The right panel shows a vertical profile cross section {\color{black} (taken in the range $655-705\ \text{nm}$)}, which corresponds to the microwave frequency resolved  ODMR signal.}}
	\label{fig_odmr_map}
\end{figure}

Fig. \ref{fig_odmr_map}. shows individual ODMR spectra combined together to form a so-called ODMR map. This closely {\color{black} resembles emission-excitation maps used in photoluminescence}, with the exception that one axis is the microwave frequency, while the other is the wavelength of the emitted light. This measurement was carried out in zero external magnetic field in the form of wavelength sweeps at fixed MW frequency. Here, the microwave sweeper is set to continuous-wave mode and wavelength resolved photoluminescence (PL) is recorded as the grating of the spectrometer is rotated. In our setup, we simultaneously detect the DC voltage (the PL signal) and the AC component (the signal detected by the lock-in amplifier) which is the ODMR signal.

The measurement time is determined by the required resolution and range in wavelength and microwave frequency and the integration time. The map presented in Fig. \ref{fig_odmr_map} was recorded with $0.5\ \text{nm}$ wavelength resolution and $100\ \text{ms}$ integration time. The microwave frequency resolution is $2\ \text{MHz}$ in the middle and was changed in bigger steps ($10-50 \text{MHz}$) moving away from the resonance resulting in $45$ different values. Altogether, this results in measurement time of $52.5$ minutes.

\subsection{$\mu$-PCD on silicon wafers}

\begin{figure}[!h]
	\centering
	\includegraphics*[width=\linewidth]{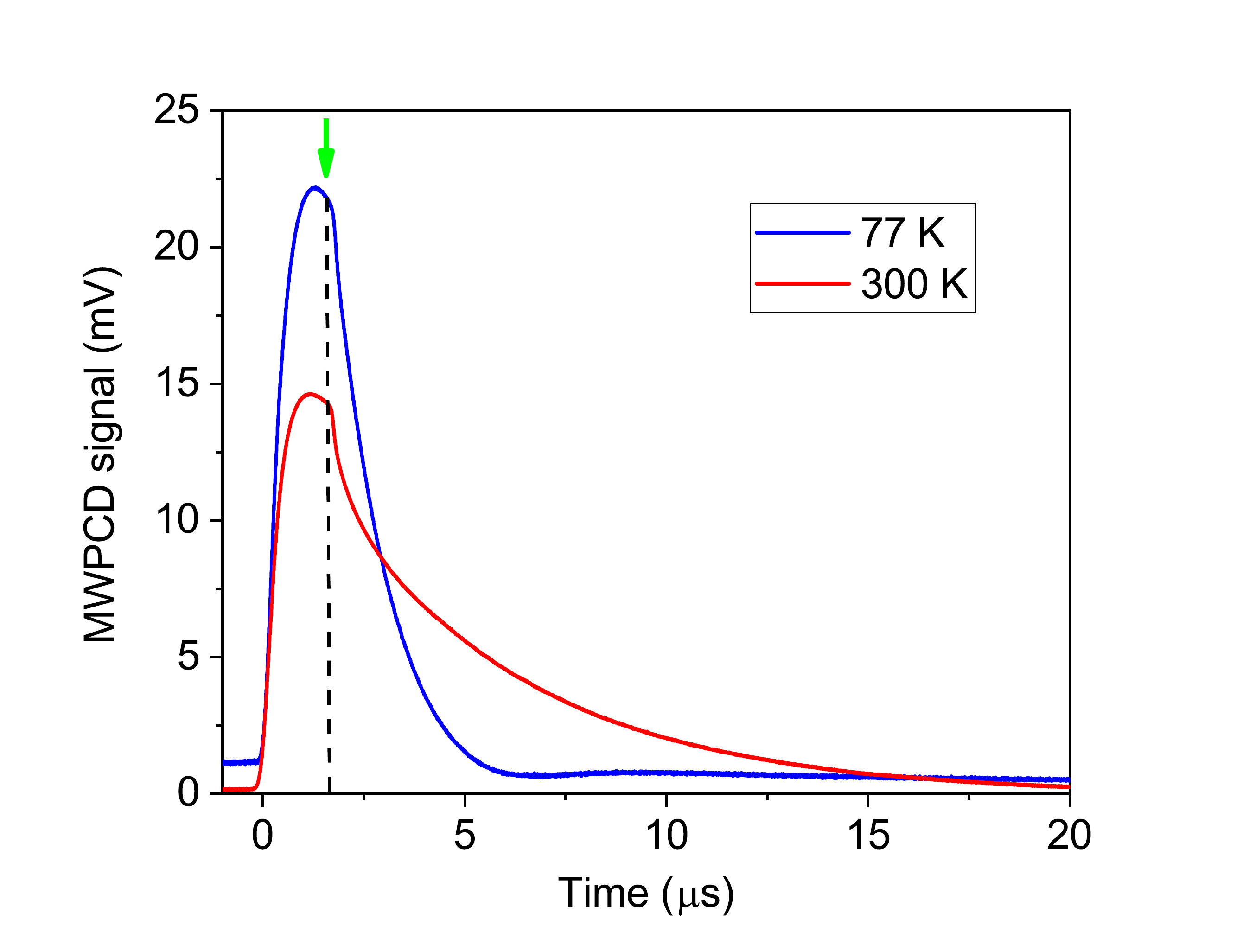}
	\caption{\emph{Microwave detected photoconductivity decay measurement carried out on phosphorus doped silicon. Red line shows the decay at room temperature and the blue spectrum was recorded at $77\ \text{K}$. The green arrow and the dashed line indicate the end of the laser pulse.}}
	\label{fig_mupcd}
\end{figure}

Fig. \ref{fig_mupcd} shows the microwave detected photoconductivity decay (or the $\mu$-PCD signal) measured on a phosphorus doped silicon sample at room temperature and at $77\ \text{K}$. The signal magnitude was obtained from the square root of the squared sum of the I and Q components. The coplanar waveguide can be readily adapted to cryostat experiments as it can be conveniently placed on the cold finger of an arbitrary cryostat. In contrast, a microwave resonator usually suffers from a change in the resonant frequency (and the quality factor) as a function of the temperature which affects the measurement sensitivity and also care is required to tune the source frequency to the of the resonator eigenfrequency. The CPW design is free from all these complications. 

In case of time resolved microwave detection, the use of microwave resonators is also impractical as the resonator bandwidth (typically 100 kHz-1 MHz) significantly limits the available time resolution \cite{Gyure}. The typical power to microwave field conversion is $2.17\cdot10^{-12}\ \text{T}^2/\text{W}$ for a TE011 type cylindrical cavity. Following the calculations presented in Ref. \cite{SIMONScalculation} for our CPW with gaps of $250\ \mu\text{m}$ separated by $1400\ \mu\text{m}$ and total width of $6\ \text{mm}$, this factor is $0.88\cdot10^{-8}\ \text{T}^2/\text{W}$. This means that the CPW performs as a cavity with a Q-factor of 4000.

As Fig. \ref{fig_mupcd} demonstrates, the cooling leads to shortening of charge carrier lifetime in the investigated material. This is described by the Shockley-Read-Hall theory \cite{Hall1952}. The impurities introduce empty gap states thus giving a rise to recombination. However, these levels can be occupied by thermally excited electrons. With decreasing temperature, fewer states are occupied therefore more recombination centers are present in the sample and the optically excited electrons have a shorter lifetime.

\section{Summary}
We presented the development of an ODMR spectrometer and a microwave photoconductivity decay measurement setup. Both instrument benefit from the coplanar waveguide structure, namely the easy optical access to the sample compared to classical microwave resonators and the broadband operation. Low temperature $\mu$-PCD measurements reveal decreasing carrier lifetime as the silicon sample is cooled. We pointed out that in specific cases resonators can outperform CPWs in terms of accuracy and sensitivity, but their advantage turns out to be a disadvantage when one needs to change measurement parameters such as microwave frequency or sample temperature. The fast $\mu$-PCD setup could reveal information on charge carrier dynamics in intensively studied novel photovoltaic materials such as methylammonium halide perovskites.

\section{Acknowledgement}
Work was supported by the Hungarian National Research, Development and Innovation Office (NKFIH) Grant Nr. K119442 and 2017-1.2.1-NKP-2017-00001.

\bibliography{pssb2020_cpw_odmr_mupcd}

%\newpage

%\section*{Graphical Table of Contents\\}
%GTOC image:

\end{document}